\documentstyle[12pt]{article}
\newcommand{\alt}{\mathrel{\raisebox{-.6ex}{$\stackrel{\textstyle<}{\sim}$}}}
\newcommand{\agt}{\mathrel{\raisebox{-.6ex}{$\stackrel{\textstyle>}{\sim}$}}}

\font\tenrm=cmr10
\font\tenit=cmti10
\font\elevenbf=cmbx10 scaled\magstep 1
\font\elevenrm=cmr10 scaled\magstep 1
\font\elevenit=cmti10 scaled\magstep 1

\font\ninerm=cmr9

\font\sevenrm=cmr7
\font\fiverm=cmr5
\font\tensym=cmsy10
\newcommand{\capfont}{\baselineskip=12pt \tenrm
\textfont0=\tenrm \scriptfont0=\sevenrm \scriptscriptfont0=\fiverm }
\newcommand{\capit}{\textfont1=\tenit}
\newcommand{\caprm}{\textfont1=\tenit}
\newcommand{\capsym}{\textfont2=\tensym}
\begin{document}
\begin{flushright}
UCLA 93/TEP/30 \\ August 1993
\end{flushright}
\begin{center}
\vglue 0.6cm
{ {\elevenbf \vglue 10pt PROSPECTS FOR MULTIPLE WEAK GAUGE BOSON \\
\vglue 3pt PRODUCTION AT SUPERCOLLIDER
ENERGIES
\footnote{
\ninerm
\baselineskip=11pt
Talk given at the {\it Workshop on Physics at Current Accelerators
and the Supercollider}, June~2-5, 1993 Argonne National Laboratory}
}
\vglue 1.0cm
{\tenrm DUNCAN A. MORRIS\footnote{email: morris@uclahep.physics.ucla.edu}\\}
\baselineskip=13pt
{\tenit Department of Physics, University of California, Los Angeles,\\}
\baselineskip=12pt
{\tenit 405 Hilgard Ave., Los Angeles, CA 90024-1547  USA}
 }
\end{center}
\vglue 0.6cm
\begin{center}{\tenrm ABSTRACT}\end{center}
\begin{center}
\parbox{5.5in}{\tenrm
We discuss the prospects for observing
multiple weak gauge boson production at the SSC and LHC.
We summarize conventional perturbative cross sections for
processes involving 1-6 final state weak gauge bosons
and compare them with more speculative scenarios
including 1) a toy model of a
strongly interacting Higgs sector patterned after hadronic
multipion production and 2)
the nonperturbative production
of \hbox{$\capfont
\,\raisebox{-.6ex}{$\capfont\stackrel{>}{\mbox{$\capsym \sim$}}$}
\,\mbox{$\capit{\cal O}$}
(\alpha_W^{-1}) \simeq 30$}
weak gauge bosons in a weakly coupled gauge sector.
}\end{center}


\vglue 0.6cm
{\elevenbf\noindent 1. Introduction}
\vglue 0.4cm
\baselineskip=14pt
\elevenrm
The purpose of the Superconducting Super Collider (SSC)
is to explore the nature of the electroweak symmetry breaking sector
and, in the process, possibly  provide a
first glimpse beyond the familiar physics of the minimal
standard model (MSM). With the realization of the
SSC on the horizon, the phenomenology of a MSM Higgs boson
between $ 80~{\rm GeV} \alt M_H \alt 800~{\rm GeV}$
has been repeatedly reworked in order to demonstrate the
feasibility of a comprehensive search$^1$.
But aside from the tangible goal of the Higgs boson
there could also be new phenomena awaiting us in the
domain of multi-TeV weak interactions.

In this paper we discuss the
phenomenology of possible
surprises in multi-TeV weak interactions at hadron supercolliders.
In particular we investigate whether multiple weak
gauge boson production can effectively
signal the presence of nonperturbative
phenomena in either the electroweak symmetry breaking sector or
the weak gauge sector. For definiteness, we concentrate
on two specific scenarios: 1) a toy model for
strong dynamics in the weak symmetry breaking sector
where the inelastic scattering of  longitudinal weak gauge
bosons  is patterned after inelastic hadronic
multipion processes$^2$ and
2) a model in which nonperturbative effects in the weak gauge sector
lead to potentially large cross sections for the production
of $ \agt {\cal O}(\alpha_W^{-1}) \simeq 30$ weak gauge bosons$^{3-5}.$
We focus on these scenarios not because there is convincing
theoretical justification for either
but rather because both admit a straightforward
analysis which suggests the prospects for their observation
(if indeed they exist at all).
Our analysis will typify the requirements
for exposing new phenomena involving multiple gauge boson production
in the environment of a hadron supercollider.

The outline of the paper is as follows. In sect.~2 we present the
conventional expectations for multiple gauge boson production
at hadron colliders. Sect.~3 discusses
 a strongly interacting Higgs sector
modeled after hadronic multipion processes and sect.~4
outlines the spectacular signatures anticipated if
nonperturbative phenomena in the weak gauge sector
appear above parton-parton center of mass \nopagebreak[3] energies in
the multi-TeV range. In sect.~5 we summarize our results and conclude.
\pagestyle{plain}

\begin{center}
\hspace*{0in}

\parbox{12cm}{ \capfont
Fig.~1. Production rates of weak gauge bosons at accelerators assuming:
${\cal L}_{Sp\bar{p}S}~=~6~\times 10^{30}~{\rm cm}^{-2}~{\rm s}^{-1}$
with two detectors at  $\sqrt{s_{p\bar{p}}}~=~.63~{\rm TeV}$;
${\cal L}_{\rm Tevatron}~=~6~\times~10^{30}~{\rm cm}^{-2}~{\rm s}^{-1}$
with two detectors at $\sqrt{s_{p\bar{p}}}~=~1.8~{\rm TeV}$;
${\cal L}_{\rm LEP}~=~1.1~\times 10^{31}~{\rm cm}^{-2}~{\rm s}^{-1}$
with four detectors at  $\sqrt{s_{e^+e^-}}~=~91.17~{\rm GeV}$;
${\cal L}_{\rm SSC}~=~10^{33}~{\rm cm}^{-2}~{\rm s}^{-1}$
with two detectors at $\sqrt{s_{pp}}~=~40~{\rm TeV}$;
${\cal L}_{\rm LHC}~=~10^{34}~{\rm cm}^{-2}~{\rm s}^{-1}$
with two detectors at $\sqrt{s_{pp}}~=~14.6~{\rm TeV}.$
Rates are dominated by processes involving single W and Z production.
Hadron collider cross sections are from Ref.~6.}
\end{center}

{\elevenbf\noindent 2. Conventional Perturbative Results}
\vglue 0.4cm
\baselineskip=14pt
\elevenrm

Since the 1983 CERN debut of the $W$ and $Z$ bosons,
progress in accelerator technology has fostered the notion
that current and future facilities effectively act as
$W$ and $Z$ factories. The SSC and LHC will take the
industrialization of weak gauge boson production
well beyond the current prolific success of LEP
and lead to the production rates of
${\cal O}(10^2-10^3)$ bosons/s (see Fig. 1).
Not only will hadron supercolliders make single
weak gauge boson production typical, but also they
will advance the study of multiple weak gauge boson processes.

\begin{center}
\hspace*{0in}

\parbox{10cm}{ \capfont
Fig.~2 Perturbative cross sections relevant to
weak gauge boson production at the SSC. Processes involving
strongly produced $t$ quarks are indicated by darker shading.
$\sigma( \mbox{$\caprm W^++W^-$} )$
and $\sigma( \mbox{$\caprm Z$} )$ are from Ref.~6. Rates for nonresonant
\mbox{$\caprm WWWW$} and \mbox{$\caprm ZZZZ$}
 production are from Ref.~8;
$\sigma(\mbox{$\caprm WWWZ$})$,
$\sigma(\mbox{$\caprm WWZZ$})$ and
$\sigma(\mbox{$\caprm WZZZ$})$
are estimated assuming
$\sigma(\mbox{$\caprm 4Z$}) =
         \mbox{$\capit f^4$}\sigma(\mbox{$\caprm 4W$})$
so that each additional \mbox{$\caprm Z$} reduces the cross section
by $\mbox{$f \capit$} \simeq  .43.$ Nonresonant $6t$ rate is estimated
assuming $\sigma(6t) / \sigma(4t) = \sigma(4t) / \sigma(2 t).$
Remaining cross sections are compiled from Ref.~7.}
\end{center}

Figure 2 indicates the relative importance of various
perturbatively calculable processes relevant to
weak gauge boson production at the SSC.
One of the most notable features is the role of
$t$ quark contributions: each $t$ quark decay in the
standard model produces a $W$ boson. As pointed out by
Barger, Stange and Phillips$^{7}$ copious $W$ production from
the weak decays of strongly produced $t$ quarks presents a
serious background to processes such as nonresonant
$WWWW$ production.

By multiplying the processes of Fig.~2
by the appropriate branching fractions to weak gauge bosons,
one obtains in Fig.~3 the SSC cross section as a function
of the total weak gauge boson multiplicity $n_W + n_Z.$
An annual SSC integrated luminosity of $10~{\rm fb}^{-1}$ suggests
the production of (but not necessarily detection of!) ${\cal O}(1)$
event/year containing five or more weak gauge bosons. This
provides us with
a useful reference point for our subsequent
discussion of more speculative
mechanisms for multiple gauge boson production.
\begin{center}
\hspace*{0in}

\parbox{14cm}{ \capfont
Fig.~3 SSC cross section for total weak gauge boson multiplicity
$n_W + n_Z$ summed over the processes of Fig.~2. Also shown (dashed)
is the cross section omitting contributions from $t$ quarks
(dark shaded processes in Fig.~2).}
\end{center}
{
\elevenbf\noindent 3. A Strongly Interacting Symmetry Breaking Sector?}
\vglue 0.4cm
\baselineskip=14pt
\elevenrm
The MSM has repeatedly defended
its position as an accurate and economical description of Nature
at all energy scales probed by experiments to date.
This is not to say
that the MSM is a definitive model; rather, we have not yet
achieved laboratory energies sufficient to test all its
attendant features --- this is the purpose of the
SSC. Aside from searching for the Higgs boson,
supercolliders can also investigate the nature of the breakdown
of $SU(2)_L \times U(1)_Y.$ Whereas the Higgs potential of
the MSM can accommodate a weakly coupled theory, it might
turn out that the breakdown of $SU(2)_L \times U(1)_Y$ is
actually
due to  underlying strong dynamics ({\it e.g.}, like
technicolor$^9$).

The obvious place to look for possible effects of underlying
strong dynamics is in the scattering
of longitudinal weak gauge bosons $W_L \sim {W^{\pm}_L,Z_L}$
which, through the equivalence theorem of Cornwall,
Levin and Tiktopoulous$^{10}$, are intimately related to
the corresponding Goldstone bosons $w \sim {w^\pm,z}$
ensuing from symmetry breakdown. The simplest manifestation
of strong dynamics would be the appearance of resonances
in assorted $w w\rightarrow w w$ channels\footnote{
For the remainder of this section we will
exploit the equivalence theorem and express our
results in terms of the equivalent Goldstone bosons $w \sim \{w^\pm,z\}.$}.
Though there is not
yet a preferred candidate for a theory of such resonances,
a variety of phenomenological models involving resonances
has been studied with moderately encouraging results$^{11}$.
In this section we wish to go beyond strong
$(2 \rightarrow 2)$ scattering in the symmetry breaking sector
and discuss the possibility of inelastic
$(2 \rightarrow~{\rm many})$ processes.
Though such channels may not be a
priority for discovering underlying
strong dynamics, they would nevertheless
be present and should be investigated.

The absence of a viable theory of strong dynamics for
$w w \rightarrow w w$ scattering is exacerbated when one
contemplates $w w \rightarrow {\rm multi-}w$ processes.
To circumvent this problem, consider the following analogy.
The symmetry-breaking Lagrangian of the MSM is that of a
$SU(2)_L \times SU(2)_R$ chirally symmetric linear $\sigma$ model:
the same type of model that successfully describes low-energy
$\pi\pi$ scattering ({\it e.g.,} for $\sqrt{s^{\pi\pi}} \alt 700~{\rm MeV}$).
The correspondence between the two theories is expressed by the
associations
\begin{eqnarray}
w & \longleftrightarrow & \pi , \\
H & \longleftrightarrow & \sigma ,\\
v \simeq 246~{\rm GeV} & \longleftrightarrow & f_\pi \simeq 93~{\rm MeV},
\end{eqnarray}
so that, at least formally, linear $\sigma$ model predictions
for low-energy $\pi \pi$ scattering are related to MSM predictions
for $w w$ scattering at a cms energy $\sqrt{s^{ww}}$ by equating

\begin{equation}
\sqrt{s^{ww}} = \displaystyle{v\over f_\pi} \sqrt{s^{\pi\pi}} \simeq
2600 \sqrt{s^{\pi\pi}}.
\end{equation}

The logical leap we propose is to take the analogy between
the low-energy limits of two models ({\it i.e.,} between
the linear $\sigma$ model below $\sqrt{s^{\pi \pi}} \alt 700~{\rm MeV}$ and
the MSM symmetry breaking sector below
$\sqrt{s^{\pi \pi}} \alt 2~{\rm TeV}$)
and, by fiat, to extend the analogy to higher energies. By doing so,
we treat $w w \rightarrow {\rm multi-}w$ processes as
 scaled-up versions of multipion production
(with the {\it same} scaling factor as in Eq.~4).
Admittedly, there is no compelling physics reason why Nature should
use inelastic pion physics
as a detailed template for $ww$ scattering --- we adopt this ansatz
simply for its definiteness. In general, if the underlying strong
dynamics does not replicate pion physics there will be additional
weak Goldstone bosons in the spectrum$^{12}$
and the scaling in Eq.~4
is modified with the consequence that inelastic multi-$w$
phenomena generally appear  below $\sqrt{s^{ww}} \simeq 2~{\rm TeV}.$

Once the decision is made to pattern
$w w \rightarrow {\rm multi-}w$
processes after multipion production, many
details follow naturally.  As one
anticipates a near-constant $\pi \pi$
cross section of $\sigma^{\pi \pi}_{\rm total}
\simeq {\cal O}(15~{\rm mb})$
above $\sqrt{s^{\pi\pi}_{\rm inelastic}} \agt
{\cal O}(1~ {\rm GeV})$
due to the production and decay of many
low-lying hadronic resonances, one might similarly expect
\begin{equation}
\hat\sigma(w w \rightarrow {\rm multi-}w) =
\hat\sigma_0^{ww}
\Theta( \hat s^{ww} - \hat{s}_0^{ww} ),
\end{equation}
where $\hat\sigma^{ww}_0 = (f_\pi/v)^2 \times
\sigma_{\rm total}^{\pi\pi}
\simeq {\cal O}(1~{\rm nb})$ and
$\sqrt{\hat s^{ww}_0} = ( v / f_\pi ) \times
 \sqrt{s^{\pi\pi}_{\rm inelastic}} \simeq
{\cal O}(2~{\rm TeV}).$
Just as the total inelastic $\pi\pi$ cross section
receives contributions from
final states with variable multiplicity, so too must
$\hat\sigma^{ww}_0$ be partitioned. In practice
one assumes a Poisson multiplicity distribution with a mean
determined by the average pion multiplicity for the corresponding
$\pi\pi$ process$^2$.
Limited transverse momentum in hadronic
reactions suggests $\langle p_T \rangle_w \simeq (v / f_\pi)
\times \langle p_T \rangle_{\rm QCD} \simeq 1~{\rm TeV}.$ The
cms energies of the SSC and LHC are actually too low for such
``limited'' $\langle p_t \rangle_w$ to be of concern --- for all
practical purposes multi-$w$ final states would be essentially isotropic.

The $pp$ cross section for multi-$w$ production
follows from Eq.~5 by using the effective vector boson
approximation$^{13}$
and writing
\begin{equation}
\sigma^{pp}_{{\rm multi-}w}(\sqrt{ s } ) =
 {\displaystyle \sum_{  w_i w_j } }\, \int \, \,
 dx_1 dx_2 \, \, f_{w_i}(x_1)
                   \, f_{w_j}(x_2) \,
\hat \sigma_0^{ww}  \, \Theta( x_1 x_2 s - \hat{s}_0^{ww} ).
\end{equation}
The double sum extends over $w_i \sim \{ w^\pm , z \} $
where $f_{w_i}(x)$ is the distribution function of $w_i$  carrying
a fraction $x$ of the original proton momentum.
Specifically,
\begin{equation}
f_{w_i}(x) =
{\displaystyle \sum_k \int_x^1
{\displaystyle dy \over y }
 f_k(y) P_{w_i/k}
\left( {\displaystyle x \over y } \right) },
\end{equation}
where $f_k(y)$ is the distribution function for quarks
or antiquarks of species $k$ inside a proton. The splitting
function $P_{w_i/k}(x)$ is the probability that a Goldstone
boson $w_i$ carries away a momentum fraction $x$ from
a parent quark of species $k.$
Figure 4 compares the ${\rm multi-}w$ cross sections at the
SSC and LHC (for $\hat\sigma_0^{ww} = 1~{\rm nb} ,
\sqrt{\hat s_0^{ww} } = 2 ~{\rm TeV}$)  to  the conventional
contributions of Fig. 3. Before any $w$ or $z$ decays are
taken into account, $\sigma^{pp}_{{\rm multi-}w}
 \simeq 33~{\rm fb}~(.85~{\rm fb})$ at the SSC (LHC).

        Practical signatures of multiple weak gauge boson
production should be phrased in terms of jets and high-$p_T$
leptons.  To do this one must impose the relevant $W$ and $Z$
branching fractions on the multi-$w$ signal,
the background processes of Fig.~2 as well as
on processes related to those of Fig.~2 by additional  QCD
bremsstrahlung. A simple  parton-level
analysis based on counting the number of jets plus leptons then
gives an idea of the signatures required to see multiple-$w$
production.

Consider multi-$w$ signatures composed of
$n_Z$ leptonically reconstructed $Z$'s, $n_l$ prompts leptons
(excluding those from $Z$ decay)
and $n_{\rm jets}$ hadronic jets.
Assuming $\hat\sigma_0^{ww} = 1~{\rm nb}$ and
$\sqrt{\hat s_0^{ww} } = 2 ~{\rm TeV}$
the SSC cross section for signatures with $n_Z=0$, $
n_{\rm jets}/2 + n_l \ge 9$ (with $n_l\ge 2,
n_{\rm jets} \ge 6)$ is $\simeq .5$~fb with a background
of $\simeq .02~{\rm fb}.$ Requiring one leptonically reconstructed
$Z$ ($n_Z=1$) and $n_{\rm jet}/2 + n_l \ge 6$
(with $n_l  \ge 1, n_{\rm jets} \ge 6$) results in a
signal (background) of $\simeq .6$~fb ($\simeq .03$)~fb.
Requiring more than one leptonically reconstructed $Z$
does not enhance the signal. In summary, the
$\simeq 33~{\rm fb}$ total yield of multi-$w$ production
is reduced to ${\cal O}(1~{\rm fb})$ of usable signal
which corresponds the production of $\agt 6$ Goldstone bosons.
Though such a simple analysis can not be considered conclusive,
it nevertheless gives a flavour of the signatures required.

\begin{center}
\hspace*{0in}

\parbox{13cm}{ \capfont
Fig.~4. Contributions to multiple weak gauge boson production
the SSC (solid) and LHC (dashed) from conventional
processes (see Fig. 2) and a strongly interacting Higgs
sector modeled after inelastic $\pi\pi$ physics. The multi-$w$
curves correspond to $\hat\sigma_0^{ww} = 1~{\rm nb},
\sqrt{\hat s_0^{ww} } = 2 ~{\rm TeV}$ in Eq. 6.}
\end{center}

{\elevenbf\noindent 4. A Nonperturbative Weak Gauge Sector?}
\vglue 0.4cm
\baselineskip=14pt
\elevenrm
Considerable excitement was generated a few years ago when
Ringwald$^3$ and Espinosa$^4$ pointed out a curious problem
involving the production of
${\cal O}(\alpha_W^{-1}) \simeq 30$ weak gauge bosons
in the multi-TeV range. The excitement (and controversy)
concerned the
results of a classical approximation to a
nonperturbative problem involving the
violation of baryon plus lepton number in
high energy standard model processes.
Whereas it was generally believed that the
amplitudes for such processes are exponentially
suppressed by a tunneling
factor  $e^{-2\pi/\alpha_W} \simeq 10^{-85},$
Ringwald and Espinosa demonstrated that enormous
combinatorial factors arising from the large number of
final state particles might come to the rescue and
compensate for the presumed suppression. Unfortunately,
the classical approximation is known to become unreliable in
precisely the region where the interesting claims
were being made; in this context it is difficult
to say whether the suggestions of
relatively unsuppressed multiple gauge
boson production should be taken
seriously or whether they are simply calculational artifacts.

        The findings of Ringwald and Espinosa inspired
considerable activity intent on sorting fact from fiction concerning
multiple gauge boson production. Cornwall$^{14}$ and
Goldberg$^{15}$ pointed
out that dreams of unsuppressed multiple gauge boson production
need not be restricted to baryon plus lepton number violating processes:
the failure of weak perturbation theory for the production
of ${\cal O}(\alpha_W^{-1}$) weak bosons is a
generic feature of large-order processes. The situation today
remains unresolved. While no one has
demonstrated the theoretical necessity
for observably large cross sections
for nonperturbative processes involving
multiple gauge boson production,
likewise no one has produced irrefutable opposing arguments.
The only consensus is that perturbation theory, when
applied to large order processes ( {\it e.g.,} like the production
of ${\cal O}(\alpha_W^{-1})$ weak bosons) breaks down
somewhere in the multi-TeV range; the SSC can test whether or
not the corresponding physical cross sections are large or small.

        Whereas the nonperturbative multi-$w$ processes
discussed in Sect.~3 were due to a strongly coupled theory
and produced only a handful of Goldstone bosons,
the nonperturbative phenomena we wish to discuss in this
section are due to the failure of perturbation theory
for the production of ${\cal O}(30)$ weak gauge bosons
in a weakly coupled theory. To distinguish between these
two cases, let us refer to the former as multi-$w$
processes (emphasizing the role of strong dynamics between
Goldstone bosons $w^\pm,z$ in the Higgs sector)
and to the latter as multi-W processes (which in principle
can act in gauge sector).

In the absence of firm theoretical guidance, we parameterize
nonperturbative parton-parton multi-W processes by
\begin{equation}
\hat \sigma(qq \rightarrow {\rm multi}-W )
= \hat \sigma_0^{WW} \Theta( \hat s^{qq} - \hat s_0^{WW} ) ,
\end{equation}
where the scale $\hat s_0^{WW}$ is set
by the breakdown of perturbation theory for processes
involving the production of ${\cal O}(\alpha^{-1}_W) \simeq 30$
weak gauge bosons ({\it i.e.,} $\sqrt{\hat s_0^{WW}} \agt 30 M_W \simeq
2.4~{\rm TeV}$). While there is no strong motivation
for the range of $\hat \sigma_0^{WW},$ an  optimistic
range to  consider might encompass

\begin{equation}
{\displaystyle \alpha_W^2 \over M_W^2 } \simeq 100~{\rm pb}
\quad \alt \quad \hat\sigma_0^{WW} \quad
\alt \quad \sigma_{\rm inelastic}^{pp} \times
\left( {\displaystyle 1~{\rm GeV} \over M_W }\right)^2 \simeq
10~\mu {\rm b},
\end{equation}
where the lower limit is characteristic of a geometrical
weak cross section and the upper limit follows from an analogy
between the weak SU(2) gauge sector and the color SU(3) gauge
sector$^5$.

\begin{center}
\hspace*{0in}

\parbox{10cm}{ \capfont
Fig.~5. Contributions to multiple weak gauge boson processes at the
SSC (solid) and LHC (dashed) from conventional perturbatively
calculable processes (see Fig. 2) and hypothetical nonperturbative
processes involving the production of ${\cal O}(\alpha_{W}^{-1})\simeq$~
30 weak gauge bosons. The multi-W curves correspond to
 $\hat\sigma_0^{WW} = 1~{\rm pb},
\sqrt{\hat s_0^{WW} } = 10~{\rm TeV}$ in Eq. 10.}
\end{center}

Folding the $qq$ cross section with quark distribution functions yields
\begin{equation}
\sigma^{pp}_{{\rm multi-}W}(\sqrt{ s } ) =
 {\displaystyle \sum_{  i j } }
 \, \int  \,
 dx_1 dx_2 \, \, f_{i}(x_1)
                   \, f_{j}(x_2)  \,
\hat \sigma_0^{WW}  \, \Theta( x_1 x_2 s - \hat{s}_0^{WW} ).
\end{equation}
Figure 5 shows the expected cross sections at the SSC and LHC
for $\hat\sigma_0^{WW} = 1~{\rm pb},$
$\sqrt{\hat{s}_o^{WW}}=10~{\rm TeV}.$
For this set of
parameters $\sigma^{pp}_{{\rm multi-}W}$
 $\simeq$~300~fb (.02~fb) at the SSC (LHC).
We interpret $\hat\sigma_0^{WW}$ as the parton-parton
cross section summed over all gauge boson multiplicities; for
purposes of illustration, Fig.~5  partitions the cross section
according to a Poisson distribution with a mean of 30.
The enormous multiplicity of central hadrons, photons and prompt
leptons from the simultaneous decays of 30 gauge bosons
has no conceivable background in the MSM. The striking nature
of such events at colliding beam facilities,
especially their central nature and large traverse energy
would make even one or two events sufficient to
demonstrate the presence of phenomena beyond that
expected from perturbation theory$^{5,16}$.

Given the spectacular nature of multi-W processes, one might
wonder if their existence might not already
be constrained by cosmic ray physics experiments.
Surprisingly, non-accelerator experiments
impose few firm constraints on such phenomena$^{17}$.
If $\sqrt{\hat s_0^{WW}} \alt 10~{\rm TeV}$ and
$\hat\sigma_0^{WW} \agt 100~{\rm nb}$ only ${\cal O}(1-1000)$
extensive air showers of multi-W origin induced by
cosmic protons would be expected in an 100~km$^2$ surface
array in one year. Unfortunately, the characteristics
of such air showers would be exceedingly difficult to distinguish from
fluctuations in a background which is ${\cal O}(10^4-10^5)$ times larger.

A more promising scenario, but one which involves an additional degree of
speculation, involves the possibility of observing
multi-W processes induced by cosmic neutrinos. Unlike
proton-induced multi-W processes which must compete
with ${\cal O}(100~{\rm mb})$ generic hadronic processes,
neutrino-induced multi-W processes compete only with
${\cal O}({\rm nb})$ generic charged current cross sections ---
but one has to first assume that a sizeable flux of ultrahigh
energy neutrinos exists!

Among the signatures of neutrino-induced
phenomena is  the underground
detection of energetic, spatially compact bundles of muons:
typical reactions occur underground so that only 2-3 prompt
muons from W decays reach the detector. Fig. 6 shows
contours in ($\sqrt{\hat s_0^{WW}},\hat\sigma_0^{WW}$) parameter space
for the expected number of horizontal
muon bundles per year in
the underwater detector DUMAND$^{18}$ assuming a cosmic neutrino flux
at the level proposed by Stecker {\it et al.}$^{19}$.
Assuming the same cosmic neutrino flux, the Fly's Eye$^{20},$
an array which is sensitive to extensive air showers,
excludes the shaded region of Fig.~6. Of course if
the required flux of ultrahigh energy neutrinos is absent
then absolutely no constraints can be put on multi-W processes
in this manner: one must then wait for the SSC and LHC.
For comparison, Fig.~6 also shows contours for 1 and 10
multi-W events in one year of operation of the SSC and LHC.

\vglue 0.6cm
{\elevenbf\noindent 5. Summary}
\vglue 0.4cm
\baselineskip=14pt
\elevenrm
It is an open question whether or not the behavior of the
electroweak symmetry breaking sector at the SSC and LHC
will be described in detail by a weakly coupled MSM.
If the Higgs sector is strongly coupled, new dynamics
may be present with the possibility of multiple weak gauge
boson production. We have examined a scenario in
which underlying strong dynamics in the Higgs sector is patterned
after hadronic multipion production and found ${\cal O}({\rm fb})$ signals
for the production of $\agt 6$ Goldstone bosons. Though such
a model is not intended to be taken literally, it typifies
the obstacles encountered if one wishes to exploit the
features of nonperturbative phenomena in the Higgs sector.

We have also reviewed the implications
of possible large cross sections for the production of ${\cal O}(30)$
weak gauge bosons due to nonperturbative phenomena
in the standard model. The SSC and LHC will be the definitive
tools for deciding whether or not the cross sections for such
processes are large.
\begin{center}
\hspace*{0in}

\parbox{10.5 cm}{ \capfont
Fig.~6. Regions of multi-W parameter space accessed
by accelerator and non-accelerator
experiments. DUMAND contours correspond to the number of
muon bundles (2-3 muons/bundle) in $10^7$~s at zenith angles greater
than $80^{\rm o}$ assuming the cosmic neutrino flux of
Stecker \mbox{\tenit et al.}$^{19}$. Shaded region is excluded by the
Fly's Eye array if the Stecker \mbox{\tenit et al.} neutrino flux is assumed.
SSC and LHC contours correspond to the number of multi-W events
expected in $10^7~{\rm s}$ of operation.}
\end{center}

\vglue 0.6cm
{\elevenbf\noindent 6. Acknowledgements}
\vglue 0.4cm
\baselineskip=14pt
\elevenrm

Many of the topics discussed in this paper are derived from
enjoyable collaborations with R.D. Peccei, A. Ringwald and R. Rosenfeld.
The local organizers of the Madison SSC Symposium
in March 1993 and the Argonne Workshop in June 1993 are to
be commended for providing a stimulating atmosphere in
which to work. \nopagebreak[3] D.A.M. is supported by the Eloisatron project.

\vglue 0.6cm
{\elevenbf\noindent 7. References}
\vglue 0.4cm
\baselineskip=14pt
\elevenrm

\begin{enumerate}
\item
SDC Technical Design Report, SDC-92-201, April 1992;
GEM Technical Design Report, GEM-TN-93-262, April 1993.

\item D.A. Morris, R.D. Peccei and R. Rosenfeld,
{\elevenit Phys. Rev.} {\elevenbf D47} (1993) 3839.

\item  A. Ringwald,
{\elevenit Nucl. Phys.} {\elevenbf B330} (1990) 1.

\item O. Espinosa,
{\elevenit Nucl. Phys.} {\elevenbf B343} (1990) 310.

\item A. Ringwald and C. Wetterich,
{\elevenit Nucl. Phys.} {\elevenbf B353} (1990) 303;
A. Ringwald, F. Schrempp, and C. Wetterich, {\elevenit ibid.}
{\elevenbf B365} (1991) 3.

\item F.T. Brandt, G.~Kramer and S.L. Nyeo,
{\elevenit Int. J. Mod. Phys.} {\elevenbf A6} (1991) 3973.

\item V. Barger, A.L. Stange, and R.J.N. Phillips,
{\elevenit Phys. Rev.} {\elevenbf D45} (1992) 1484.

\item V. Barger, T. Han and H. Pi,
{\elevenit Phys. Rev.} {\elevenbf D41} (1990) 824.

\item E. Farhi and L. Susskind,
{\elevenit Phys. Rep.} {\elevenbf 74} (1981) 277.

\item J.M. Cornwall, D.N. Levin and
G. Tiktopoulos,
{\elevenit Phys. Rev.} {\elevenbf D10} (1974) 1145;
{\elevenbf 11}, (1975) 972(E).

\item See {\it e.g.}, M.S. Chanowitz, in {\elevenit Perspectives
on Higgs Physics}, edited by G. Kane (World Scientific, Singapore,
{\elevenit in press}); J. Bagger {\it et al.}, Fermilab preprint
FERMILAB-Pub-93/040-T, June 1993.

\item R.S. Chivukula, M.J. Dugan, and
M. Golden, {\elevenit Phys. Rev.} {\elevenbf D47} (1993) 2930.

\item G.L. Kane, W.W. Repko, and W.B. Rolnick,
{\elevenit Phys. Lett.} {\elevenbf 148B} (1984) 367;
S. Dawson, {\elevenit Nucl. Phys.} {\elevenbf B249} (1985) 42.

\item J.M. Cornwall,
{\elevenit Phys. Lett.} {\elevenbf B243} (1990) 27.

\item H. Goldberg,
{\elevenit Phys. Rev. Lett.} {\elevenbf 66} (1991) 1267;
{\elevenit Phys. Lett.} {\elevenbf B246} (1990) 149.

\item
G. Farrar and R. Meng,
{\elevenit Phys. Rev. Lett.} {\elevenbf 65} (1990) 3377.

\item
D.A. Morris and A. Ringwald, CERN preprint CERN-TH.6822/93, August 1993.

\item
C.M. Alexander {\elevenit et al.}, in
{\elevenit  Proc. 23rd Int. Cosmic Ray Conf.}, Calgary, July 1993,
eds. J. Wdowczyk {\it et al.}, (U. of Calgargy, Calgary, 1993) Vol. 4.,
p. 515.

\item
F. Stecker, C. Done, M. Salamon, and P. Sommers,
{\elevenit Phys. Rev. Lett.} {\elevenbf 66} (1991) 2691;
{\elevenit ibid.} {\elevenbf 69} (1992) 2691(E).

\item
C.M. Alexander {\elevenit et al.} (DUMAND Collaboration), in
{\elevenit  Proc. 23rd Int. Cosmic Ray Conf.}, Calgary, July 1993,
eds. J. Wdowczyk {\it et al.}, (U. of Calgargy, Calgary, 1993) Vol. 4.,
p. 515.

\item
R. Baltrusaitis {\it et al.} (Fly's Eye Collaboration),
{\elevenit Phys. Rev.} {\elevenbf D31} (1985) 2192.

\end{enumerate}
\end{document}